# Magnon contribution to the magnetoresistance of iron nanowires deposited using pulsed electrodeposition

**Philip Sergelius**[*,1], **Josep M. Montero Moreno**[1], **Martin Waleczek**[1], **Tim Böhnert**[2], **Detlef Görlitz**[1] and **Kornelius Nielsch**[1]

[1] Institute of Nanostructure and Solid State Physics, Universität Hamburg, Jungiusstraße 11, 20355 Hamburg, Germany
[2] International Iberian Nanotechnology Laboratory, Av. Mestre José Veiga, s/n, 4715-330, Braga, Portugal

* Corresponding author: e-mail Philip.Sergelius@physnet.uni-hamburg.de, Phone: +49 40 42838 5248, Fax: +49 40 42838 3589

Iron nanowires with a square cross section are grown by pulsed electrodeposition within a newly developed nanochannel template that allows for easy characterization. Measurements of the magnetoresistance as a function of magnetic field and temperature are performed within a large parameter window allowing for the investigation of the magnonic contribution to the magnetoresistance of electrodeposited iron nanowires. Values for the temperature dependent magnon stiffness $D(T)$ are extracted:
$$D(T) = D_0(1 - d_1 T^2) = 365 \cdot (1 - 4{,}4 \cdot 10^{-6} \cdot T^2 \cdot \text{K}^{-2}) \text{meV Å}^2$$

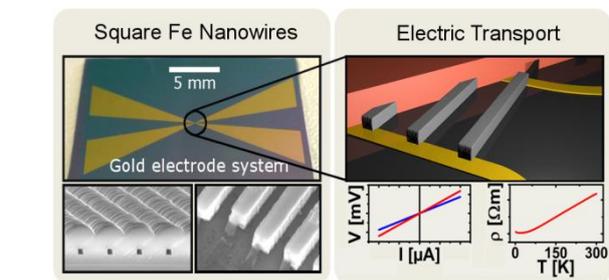

Iron nanowires reveal a high magnetization that can be combined with large shape anisotropies and very high coercive fields. These features make iron nanowires particularly interesting for applications within MRAM data storage [1,2]. There have been numerous investigations on the magnetic behavior of pure iron nanowire arrays [3] and it's alloys [4-7], but only very few that deal with the electric properties of electrodeposited nanowires [8]. This is mostly due to the fact that iron is oxidized without forming a protective native oxide as soon as it is exposed to ambient air which makes electric characterization of the unaltered material impossible. The developed template presented in this publication allows us to carry out in-situ transport measurements within the same thick oxide template that is also used for the growth of the wires. Thus, the wires are completely protected against oxidation and a good ohmic contact between conduction leads and the nanowire is guaranteed. The electric transport properties of all ferromagnets and especially ferromagnetic nanowires are dependent on the applied external magnetic field. The most prominent characteristic is the anisotropic magnetoresistance (AMR) effect, which was explained by Smit [9], suggesting an angular dependence of the scattering cross section between orbitals of atoms in ferromagnetic order and conduction electrons. As soon as all magnetic moments are fully saturated, the Lorentz force is responsible for an increase in the resistance as the magnetic field increases, since it forces electrons on a curved trajectory and thus reduces the projected free wavelength in the direction of the k-vector. Above approximately $T_c/5$ however, ferromagnets show an inverse and near linear behavior. This is due to a decrease of the total magnon population with an increasing magnetic field and thus electron-magnon scattering is suppressed. This effect is significantly stronger than the counteracting Lorentz-dependence [10].

The developed template system (Figure 1) is based on a $Si/SiO_2$-wafer onto which conduction leads are evaporated. On top of these, a nanochannel template is fabricated using a sacrificial polymer layer that is patterned with laser-interference-lithography [11,12], reactive ion etching and atomic layer deposition (ALD). The sacrificial polymer layer is calcinated at 350 °C and a hollow channel structure remains on top of the previously deposited conduction leads. The height and width of the perfectly rectangular nanochannels can be tuned within a large window down to as to as small as 40x40 nm². A hole in the nanochannel system is fabricated in the desired location by wet, dry, or focused ion beam etching which allows the electrolyte solution to enter the nanochannel system as soon as the template is submerged. By electrically contacting one of the conduction leads, electrodeposition within the nanochannel system can be triggered. The setup can be designed so that the wire growth takes place from one electrode towards another. When the growth has reached the opposite side, the deposition process can be stopped, the large pads of the gold contact leads connected and the sample can directly be mounted into the measurement setup. In-depth discussion of the fabrication steps can be found in Reference [13].

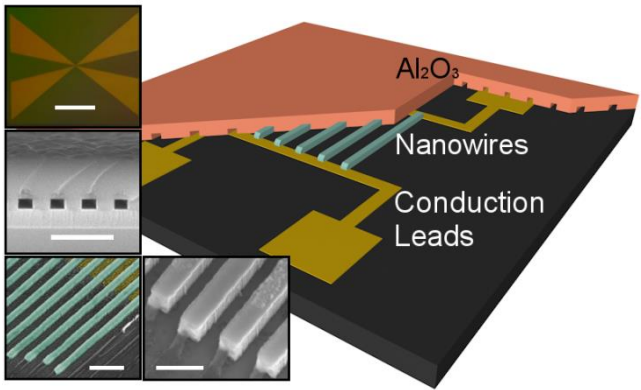

**Figure 1** Sketch of the developed template system with severely distorted scales for easier understanding. The contact pads (top inset, photo) can be designed on a cm-scale while the distance between both contacts (the area of wire growth) is approximately 10-20 μm. The second inset shows a cross-section of the hollow nanochannel template (SEM). The bottom two insets show small and large nanowires with square cross-sections (SEM). Scale bars correspond to 1 cm (top) and 400 nm (bottom three).

For the growth of iron nanowires by pulsed electrodeposition [14] an iron sulfate bath is used consisting of boric acid (4.5 g/l) as a buffer, glycine (12.2 g/l) and ascorbic acid (1.0 g/l) as antioxidating agents and iron(II) sulfate heptahydrate (4.5 g/l) used as iron source. The electrodeposition process is controlled via the potential. The deposition pulse is square with a height of -1.3 V vs. an Ag/AgCl/KCl$_{sat}$ reference electrode and a length of 10 ms. In order to allow the ion concentration to regenerate and to achieve a more stable pH-Value within the pores, ultimately resulting in more homogenous wires, the deposition pulse is followed by an off-time of 100 ms at -0.45 V respectively. The applied potential is too low to reduce Fe(II)-Ions at the cathode, but high enough to protect the material at the nanowire's growth front from being oxidized and dissolved, thus no current will flow during the off-time. The growth rate of the wires is determined experimentally and the deposition is then conducted until the wires have reached the second electrode with an ample buffer time.

The conduction leads are connected to a chip-carrier of a *PPMS Dynacool* cryostat (Quantum Design). SEM characterization reveals that three parallel nanowires with dimensions of 75x70x12000 nm³ are electrically contacted. Linear IV-curves over a wide temperature range (2 K to 300 K) and a linear R(T)-curve above 50 K confirms the metallic nature of our samples and the good ohmic contact between nanowires and conduction leads. The room-temperature resistivity of $\rho = 235\ \mu\Omega \cdot cm$ is larger than the bulk value, a behavior which is attributed to size-effects and frequently reported for other ferromagnetic nanowires [15-17]. Measurements of the magnetoresistance in the range of $-9\ T \leq B \leq 9\ T$ with B ⊥ I reveal a distinct AMR peak and a near-linear drop in resistivity for B > 3 T and for temperatures above 200 K (Figure 2).

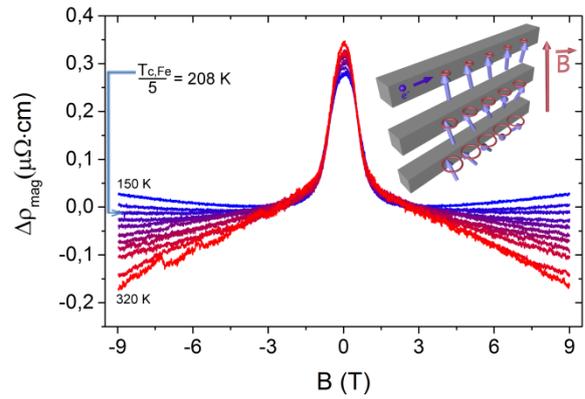

**Figure 2** Magnon magnetoresistance of iron nanowires. The crossover from a positive to a negative dependence takes place at a fifth of the curie temperature of iron.

This drop was first explained by Mott in 1964 [18] as spin-flip scattering of conduction electrons with magnons, the quasi particle corresponding to spin-waves. Higher magnetic fields or lower temperatures decrease the precession amplitude of spin-waves as shown in the cartoon in Figure 2, phenomenologically leading to a decrease of the total amount of magnons in the system, consequently reducing the electric resistance. The magnon dispersion relation can be written as

$$E(\vec{k}) = D(T)\vec{k}^2 + g\mu_{Bohr}(B + \mu_0 M),$$

with $D(T)$, the temperature dependent proportionality factor between energy and squared wave vector. This factor is often called magnon stiffness or magnon mass renormalization and its first order approximation is given by $D(T) \approx D_0(1 + d_1 T^2)$ [19,20]. In 1968, Stringfellow measured the magnon stiffness of iron whiskers using neutron scattering and reported values of $D_0 = 314\ meV\ Å^2$ and $d_1 = 5 \cdot 10^{-6} K^{-2}$ [21].

Another method for measuring these coefficients was developed by Raquet *et al.* in 2002, reporting values of $D_0 = 350\ meV\ Å^2$ and $d_1 = 2{,}5 \cdot 10^{-6} \cdot K^{-2}$ [10,22].

Using their simplified expression which is valid for fields below 100 T and in the region between $T_c/5$ and $T_c/2$, the high-field MR can be described by

$$\Delta\rho(T,B) \propto \frac{BT}{D(T)^2}\ln\left(\frac{\mu_{Bohr}B}{kT}\right). \qquad (1)$$

Since $\lim_{B\to 0} B \cdot \ln(B) = 0$, an offset $\rho_0$ is incorporated into the fitting formula to account for the fact that the measured curves do not cross the origin, resulting in the expression for the magnetic-field dependency of the resistivity

$$\Delta\rho_{mag}(T_i,B) = \rho_0 + \frac{BT_i}{D(T_i)^2}\ln\left(\frac{\mu_{Bohr}B}{kT_i}\right). \qquad (2)$$

By fitting Formula 2 to the measured data displayed in Figure 2 for each temperature $T_i$ and for fields above 3 T, we infer the temperature dependent magnon stiffness for electrodeposited iron nanowires (Figure 3).

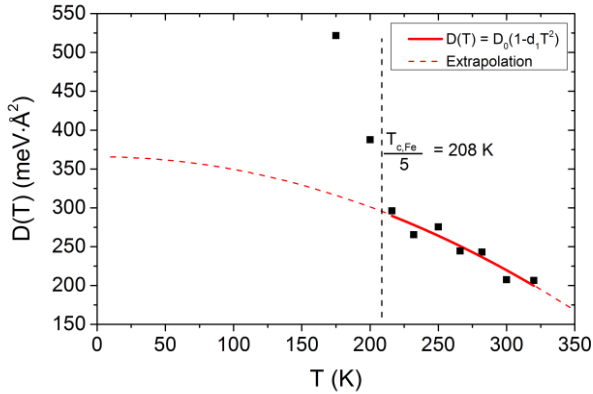

**Figure 3** Values for $D(T)$ derived from fits of Eq. (2) to data displayed in Fig (2). The dashed red line is the extrapolation towards $T=0$ K.

A fit of the acquired data for $D(T)$ for T > 200K to the first order temperature dependence of the magnon stiffness yields the following values:

$$D(T) = D_0(1 - d_1 T^2) = (365 \pm 17) \cdot [1 - (4{,}4 \pm 0{,}6) \cdot 10^{-6} \cdot T^2 \cdot K^{-2}]\,\text{meV Å}^2$$

Remarkably, the expression of Raquet (Eq. 2) is capable of describing the measurement data until exactly $T_c/5 = 208$ K which is the temperature when the slope turns from a negative to a positive dependence as depicted by the blue arrow in Figure 2 and the dashed line in Figure 3. The extrapolated stiffness $D(T=0)$ is in good agreement with values reported by Raquet *et al.*, however the difference in $d_1$ suggests a stronger temperature dependence.

In contrast to previous studies which investigate the effect of confinement in one dimension, such as in thin films [19], we aim to investigate the confinement in two dimensions. The limit is a nanowire with a 1D conduction channel. Due to the high accordance with bulk values, we conclude that the magnon stiffness is not affected by 2D-confinement until at least 70x75 nm².

In summary we have measured the magnon stiffness of electrodeposited iron nanowires using a versatile synthesis technique. This technique can of course be adapted for any other kind of nanowire that can be deposited using electrodeposition, for example Bismuth wires which are known to create a very strong surface oxide [23].

**Acknowledgements** The research leading to these results has received funding from the European Unions's 7th Framework Programme under grant agreement n°309589 (M3d) and we gratefully acknowledge financial support from the German Research Foundation (DFG) via SFB 986 "M3", project C3 and by the German Priority Program SPP 1536 "SpinCAT".